\documentclass[onecolumn]{aa}

\usepackage{graphicx,natbib}
\bibpunct{(}{)}{;}{a}{}{,}

\begin{document}

\sloppypar

\title{Reconsidering the galactic coordinate system}

\author{Jia-Cheng Liu \inst{1} \and Zi Zhu \inst{1,2} \and Hong Zhang \inst{1,2}}
\institute{Department of Astronomy, Nanjing University, Nanjing
210093, China \and Key Laboratory of Modern Astronomy and
Astrophysics (Nanjing University), Ministry of Education, Nanjing 210093, China\\
 \email{[jcliu;\,zhuzi;\,zhangh]@\,nju.edu.cn} }
\date{Received / Accepted}
\titlerunning{Reconsidering the galactic coordinate system}
\authorrunning{Liu et al.}

\abstract{Initially defined by the IAU in 1958, the galactic
coordinate system was thereafter in 1984 transformed from the
B1950.0 FK4-based system to the J2000.0 FK5-based system. In 1994,
the IAU recommended that the dynamical reference system FK5 be
replaced by the ICRS, which is a kinematical non-rotating system
defined by a set of remote radio sources. However the definition of
the galactic coordinate system was not updated. We consider that the
present galactic coordinates may be problematic due to the
unrigorous transformation method from the FK4 to the FK5, and due to
the non-inertiality of the FK5 system with respect to the ICRS. This
has led to some confusions in applications of the galactic
coordinates. We tried to find the transformation matrix in the
framework of the ICRS after carefully investigating the definition
of the galactic coordinate system and transformation procedures,
however we could not find a satisfactory galactic coordinate system
that is connected steadily to the ICRS. To avoid unnecessary
misunderstandings, we suggest to re-consider the definition of the
galactic coordinate system which should be directly connected with
the ICRS for high precise observation at micro-arcsecond level.

   }

   \maketitle

%

%
%

\section{Introduction}           
\label{sect:intro}

For studies of the galactic structure, kinematics and dynamics, it
is important to establish the galactic coordinate system having its
equatorial plane in the galactic plane, and the pole coincide with
the galactic pole. The definition of the system of galactic
coordinates, which is based on the FK4 system at the epoch of
B1950.0, was announced by the International Astronomical Union
Sub-Commission 33b in the late 1950s \citep{paper1}. It is a
heliocentric rectangular system with three principal axes in the
direction of galactic center, galactic rotation, and north galactic
pole. Following the recommendation of the Sixteenth General Assembly
of the IAU at Grenoble in 1976, the previous fundamental reference
system FK4 have been replaced with FK5 at the epoch of J2000.0. In
addition, the IAU adopted the new system of astronomical constants
and time unit in order to construct an inertial framework from
observed materials. The transformation procedures from the FK4 based
reference system to the FK5 reference system have been developed by
\citet{standish82} and \citet{aoki83}. The transformation matrix
from the equatorial coordinate system to galactic coordinate system
at the epoch of J2000.0 has been derived by \citet{murray89} under
the assumption that the FK4 and FK5 systems have only orientation
offsets due to the precession motion from B1950.0 to J2000.0 as well
as the equinox bias.

The fundamental celestial reference system for astronomical
application is now the International Celestial Reference System
(ICRS) as specified in the IAU Res. B2 of 1997. To establish the
ICRS as a practical system, the IAU provided a set of distant
benchmark radio sources, by whose coordinates the direction of the
ICRS axes are defined \citep{ma98}. The ICRS is realized optically,
but at lower accuracy, by the star positions and proper motions of
the \emph{Hipparcos Catalogue} (ESA 1997). However, the galactic
coordinate system is still based on the observations in the FK4
system at the epoch of B1950.0 or more recently, transformed to the
FK5 system in a loose sense. The FK5 system is non-inertial based on
the comparison  between the FK5 and \emph{Hipparcos} proper motion
systems. In other words, the galactic coordinate system, whether its
principal axes are in the FK4 system at B1950.0 or in the FK5 system
at J2000.0, is rotating with respect to the ICRS. This can lead to
some confusions or misunderstandings. Considering a
sub-milli-arcsecond or even micro-arcsecond astrometric accuracy,
the galactic coordinate system should be directly connected to the
ICRS for uniform practical applications and exact definition of
coordinate system. We suggest that a new definition of the galactic
coordinate system be considered.

In \S 2 and \S 3 of the present work we review the definition of the
galactic coordinate system as well as the confusion in astronomical
applications. In \S 4 a suggestion of the new definition is
presented and some conclusions are summarized in \S 5.

\section{History of the development of the galactic coordinate system}
\label{sect:history}

\subsection{IAU1958 galactic coordinate system}
From the time of William Herschel in the 1780s until 1932 various
galactic coordinate systems which differ from one another by one or
two degrees were used. \citet{ohl32} summarized extensive
observations at Harvard Observatory and adopted a position of the
north galactic pole at right ascension
$\alpha_{B1900.0}^{\,p}=12^h40^m$, declination
$\delta_{B1900.0}^{\,p}=+28^{\circ}$. With the development of the
radio astronomy in the 1950s, it became evident that the old
galactic coordinate system did not serve its purpose any longer and
was seriously in error for many observations especially in the radio
specturm. In 1958 the IAU Sub-Commission authorized a new definition
of the galactic coordinate system, according to which the equatorial
plane was defined by the flat HI principal plane and the zero point
of the system by the direction of the galactic center. Radio
continuum observations gave strong support to a pole based on the HI
observations and optical data were used for a checking purpose. The
Sun is effectively defined to be in the plane even though it is
thought to be some 10-20\,pc above the HI principal plane
\citep{zhu09}. The revision and reconstruction of the galactic
coordinate system was presented in a series of five papers
\citep{paper1,paper2,paper3,paper4,paper5}. The system was defined
by three principal values:
\begin{eqnarray}
&&\alpha_{B1950.0}^{\,p}=12^h49^m   \nonumber  \\
&&\delta_{B1950.0}^{\,p}=+27^\circ.4  \\
&&\theta_{B1950.0}=123^\circ ,    \nonumber
\end{eqnarray}
where $\alpha_{B1950.0}^{\,p}$, $\delta_{B1950.0}^{\,p}$ are the
right ascension and declination of the north galactic pole and
$\theta_{B1950.0}$ is the position angle of the galactic center from
the equatorial pole. The three principal values defining the
galactic coordinate system are based on the FK3 system because at
that time in 1958 the FK4 catalogue was not published yet.
Nevertheless, the equinox and equator were not changed in the FK4
with respect to the FK3, so it should be justified to assume that
the defining values are based on the equator and equinox of the FK4
system at the epoch of B1950.0. These numbers in Equation (1) should
be used as a strict definition with infinite accuracy, or in other
words, valid to an infinite number of decimals. Users are supposed
to quote the definition correctly rather than copy computer
algorithms in order to avoid inconsistency in the forward-backward
conversion between equatorial $(\alpha, \, \delta)$ and galactic
$(\ell, \, b)$ coordinate systems \citep{lane79}.

Note that the establishment of the galactic coordinate system does
not require the same degree of precision as the equatorial reference
system which are realized by the fundamental catalogues (FK4, FK5),
because it is constructed based on the concentration of HI gas in
statistical sense. Although the requirements for the establishment
of the galactic coordinate system are less drastic, it must be fixed
in the fundamental reference system of the time, say, aligned with
FK4, FK5, or ICRS.

\subsection{Principal galactic axes in the FK5 system at J2000.0}
The introduction of the new fundamental stellar catalogue FK5 and
the IAU(1976) system of astronomical constants were brought into
effect from 1 January 1984. The transformation from the FK4 to the
FK5 includes the following changes:
\begin{itemize}
\item the E-term of aberration was removed from the mean places
and proper motions;
\item the Newcomb's precession constants were replaced by the IAU 1976
astronomical constants \citep{lieske77};
\item the bias between FK4 equinox and the dynamical equinox (equinox
correction) was eliminated;
\item the time unit was changed from tropical to Julian centuries;
\item the standard coordinate system was precessed from the equinox
and equator at B1950.0 to that of J2000;
\item the positions of the stars were presented at the epoch of J2000.0
by adding their proper motions.
\item the systematic differences $(FK5-FK4)$ were corrected, if known.
\end{itemize}
The so-called E-term of aberration depends explicitly on the
eccentricity of the Earth orbit around the Sun and represents the
components of the displacement due to the departure of the elliptic
orbital motion from a circle. During the process of transforming the
galactic coordinates from the FK4 system (B1950.0) to the FK5 system
(J2000.0), the impure rotation related to the E-term of aberration
have consequently lead to an unacceptable non-orthogonal system of
axes if those axes are considered to be defined by specific objects
in the three principal directions \citep{murray89}. Accordingly
Murray proposed that the axes of IAU1958 galactic coordinate system
be considered as absolute directions, unaffected by the E-term of
aberration.

In order to understand the transformation, we consider the column
unit vector $(3\times1)$ of a certain absolute direction or a
fictitious distant extragalactic source at the epoch of B1950.0,
${\bf{A}}_B$ and ${\bf{G}}_B$, and at the epoch of J2000.0,
${\bf{A}}_J$ and ${\bf{G}}_J$. For brevity, the subscript $B$ stands
for the epoch of B1950.0 and $J$ stands for J2000.0. The characters
$\bf A$ and $\bf G$ denote the unit vectors in the equatorial and
galactic systems thus they can be expressed as :
\begin{eqnarray*}
{\bf{A}}_{B}&=&\left(\cos \alpha_{B} \cos \delta_{B}\;\;
\sin \alpha_{B} \cos \delta_{B}\;\; \sin \delta_{B}\right)^T \\
{\bf{A}}_{J}&=&\left(\cos \alpha_{J} \cos \delta_{J}\;\;\;
\sin \alpha_{J} \cos \delta_{J}\;\;\; \sin \delta_{J}\right)^T  \\
{\bf{G}}_{B}&=&\left(\cos \ell_{B} \cos b_{B}\;\;\;
\sin \ell_{B} \cos b_{B}\;\;\; \sin b_{B}\right)^T \\
{\bf{G}}_{J}&=&\left(\cos \ell_{J} \cos b_{J}\;\;\;\; \sin \ell_{J}
\cos b_{J}\;\;\;\; \sin b_{J}\right)^T .
\end{eqnarray*}
Let $\mathcal N_{B}$ and $\mathcal N_{J}$ be the $3\times3$ matrixes
that convert the position vector from the equatorial system to the
galactic system, i.e.
\begin{eqnarray}
 {\bf{G}}_{B} &=& \mathcal N_{B}\, {\bf{A}}_{B} \\
 {\bf{G}}_{J} &=& \mathcal N_{J}\, {\bf{A}}_{J}.
\end{eqnarray}
Suppose that $\mathcal X(t)$ is the transformation matrix of
position vector from J2000.0 system to that of B1950.0 system at
some epoch of $t$, $\mathcal X(t)$ can be computed from the
following formula:
\begin{equation}
\mathcal X(t)=\mathcal P_{NEWC}(t,B1950.0)\mathcal R_3(E(t))\mathcal
P_{IAU76}(J2000.0,t)
\end{equation}
where $t$ is reckoned from B1950.0 in tropical centuries. $\mathcal
P_{IAU76}(t_1,t_2)=\mathcal{R}_3(-z_A)\mathcal{R}_2(\theta_A)\mathcal{R}_3
(-\zeta_A)$ is the precession matrix from the epoch of $t_1$ to
$t_2$ based on IAU (1976) constant system and $\mathcal
P_{NEWC}(t_1,t_2)$ the precession matrix based on Newcomb's
precession formulae. The precession matrices were constructed for
the transformation of coordinates from one epoch ($t_1$) to another
($t_2$). For the equinox correction, \citet{fricke82} pointed out
that the equinox correction $E(t)$ at any epoch can be expressed as
\begin{equation}
E(t)=E_{1950}+\dot{E}t.
\end{equation}
We assume here that $E=0''.525$ at the epoch of J1950.0 and
$\dot{E}=1''.275/\rm{Julian \;century}$ as \citet{aoki83} did for
the consistency of the new formula for the sidereal time.

According to \citet{soma90}'s method, the E-term in aberration and
the systematic and individual corrections are ignored, thus the
equatorial position is only affected by the precession motion from
B1950.0 to J2000.0 and the equinox correction:
\begin{equation}
{\bf{A}}_{J}=\mathcal{X}_0^{-1}{\bf{A}}_{B}
\end{equation}
where the matrix $\mathcal{X}_0$ is the constant term in the
expanded polynomial of $\mathcal X(t)$:
\begin{displaymath}
\mathcal X(t)=\mathcal X_0+\mathcal X_1 t+\mathcal X_2 t^2+\cdots.
\end{displaymath}
We can see from Equation (6) that in current problem the
transformation is unrelated to the epoch $t$.

Because the galactic coordinate system must be regarded absolute in
the celestial sphere, the galactic coordinates $(\ell, b)$ of a
given direction (or a distant radio source) will not change although
its equatorial coordinates are referred to the FK4 system and then
transformed to the FK5 system. The transformation only takes the
orientation difference between the FK4 and the FK5 into account, as
if they were both inertial. Hence we have
\begin{equation}
{\bf G}_{J}={\bf G}_{B}.
\end{equation}
Finally we obtain from Equation (3), (6) and (7) the transformation
matrix based on the FK5 system at the epoch of J2000.0:
\begin{equation}
\mathcal N_{J}=\mathcal N_{B}\mathcal X_0.
\end{equation}
The numerical matrices, printed to 0.1 milli-arcsecond accuracy of
the precession formulae in the IAU (1976) system, are
\begin{displaymath}
\mathcal N_{B}= \left(
\begin{array}{rrr}
-0.066988739410 &    -0.872755765850 &    -0.483538914637  \\
+0.492728466081 &    -0.450346958020 &    +0.744584633279  \\
-0.867600811149 &    -0.188374601732 &    +0.460199784785
\end{array}
\right);
\end{displaymath}
\begin{displaymath}
\mathcal X_{0}= \left(
\begin{array}{rrr}
+0.999925679496 &    +0.011181483239 &    +0.004859003772  \\
-0.011181483221 &    +0.999937484893 &    -0.000027170294  \\
-0.004859003815 &    -0.000027162595 &    +0.999988194602
\end{array}
\right),
\end{displaymath}
and
\begin{equation}
\mathcal N_{J}= \left(
\begin{array}{rrr}
-0.054875539390 &    -0.873437104725 &    -0.483834991775  \\
+0.494109453633 &    -0.444829594298 &    +0.746982248696  \\
-0.867666135681 &    -0.198076389622 &    +0.455983794523
\end{array}
\right).
\end{equation}
Such a procedure guarantees that the galactic coordinate system is
orthogonal and non-rotating with respect to the FK5 system at the
epoch of J2000.0.

Inserting $(b=90^\circ)$ and $(\ell=0,\;b=0)$, respectively, in
Equation (3), we can obtain the direction of the galactic pole and
the position angle in the J2000.0 equatorial system (also at 0.1
milli-arcsecond level):
\begin{eqnarray}
&&\alpha_{J}^{\,p}  = \;\;12\,^h51\,^m26\,^s.27549  \nonumber  \\
&&\delta_{J}^{\,p}  \,=  +27\,^\circ07\,'41\,''.7043  \\
&&\theta_{J}  = \;\;122\,^\circ.93191857 ,   \nonumber
\end{eqnarray}
and the equatorial coordinates of the galactic center:
\begin{eqnarray}
&&\alpha_{J}^0  = \;\;17\,^h45\,^m37\,^s.19910  \nonumber  \\
&&\delta_{J}^{\,0}  \,=  -28\,^\circ56\,'10\,''.2207 .
\end{eqnarray}
\citet{murray89} and \citet{miyamoto93} have also obtained
coincidental results using a slightly different method with only a
few tenth of milli-arcsecond discrepancy. It is worth noting that if
we set $t=0$ (i.e. transformation at the epoch of B1950.0) in
Equation (4) and its expanded form, the transformation matrix is
given by :
\begin{equation}
\mathcal X_0=\mathcal R_3(E(B1950.0)) \mathcal
P_{IAU76}(J2000.0,B1950.0),
\end{equation}
which was adopted by \citet{murray89}.

The transformation between equatorial and galactic coordinate system
in the \emph{Hipparcos} system \citep{esa97} adopted Murray's method
during the conversion from the equatorial to galactic coordinates in
which the difference between the FK5 and the \emph{Hipparcos} is
ignored. It is worth noting that the \emph{Hipparcos} transformation
matrix is numerically different from ours at the milli-arcsecond
level because it is calculated by adopting $\alpha_{J}^{\,p}$,
$\delta_{J}^{\,p}$ and $\theta_{J}$ as definitional values. There is
possible loss of precision during the computing procedure, as
adopted by \emph{Hipparcos}, from the truncated value of
$\alpha^{\,p}_J$, $\delta^{\,p}_J$, $\theta_J$ to the matrix
$\mathcal N$:
\begin{displaymath}
\mathcal N_{Hip}= \left(
\begin{array}{rrr}
-0.0548755604 &    -0.8734370902 &    -0.4838350155  \\
+0.4941094279 &    -0.4448296300 &    +0.7469822445  \\
-0.8676661490 &    -0.1980763734 &    +0.4559837762
\end{array}
\right).
\end{displaymath}

\section{Confusions and misunderstandings in applications of the galactic coordinate system}
\label{sect:confuion}

As mentioned in last sections, the present galactic coordinates may
be problematic due to the inaccurate transformation from one
fundamental reference system to another. Consequently, some
confusion in applications of galactic coordinates and
misunderstandings in the concept of the reference system can be
found.

\subsection{Conversion of galactic coordinates from the FK4 to the FK5 system}
According to the IAU resolutions adopted in 1976 (Trans. IAU 16B),
in 1979 (Trans. IAU 17B), and in 1982 (Trans. IAU 18B), new methods
and astronomical constants system are used for the computation of
the positions and proper motions of stars. Conversion processes of
mean positions and proper motions of stars from the FK4 system
(B1950.0) to the FK5 system (J2000.0) had been proposed by
\citet{standish82}, \citet{aoki83} and subsequently reviewed by
\citet{smith89}, \citet{yall89}, \citet{murray89} and
\citet{soma90}. The methods for calculating the apparent places in
the new IAU system were also developed by \citet{lederle84}.

The rigorous method of converting the position and proper motion of
a celestial object at B1950.0 to those at J2000.0 contains two major
steps: (i) remove the E-terms of aberration and their variations'
effect; (ii) transfer the position and proper motion vector using a
$6\times 6$ matrix to the new IAU system. Adopting the notations
given by \citet{standish82}, the 6-space transformation matrix
operated on 1984 January 1.0 \citep{aoki83} is described as
\begin{eqnarray}
\mathcal{M}_{1984.0}=\left(
\begin{array}{cc}
I & G_2 I \\
O & I
\end{array}
\right) \mathcal{P}_{IAU76}^{-1}(J2000.0,1984.0,1)
\mathcal{Q}_3 (E(1984.0), \dot{E}) \nonumber  \\
\times \left(
\begin{array}{cc}
I & O  \\
O & FI
\end{array}
\right) \mathcal{P}_{NEWC}(B1950.0,1984.0,1) \left(
\begin{array}{cc}
I & G_1 I  \\
O & I
\end{array}
\right),
\end{eqnarray}
where $G_1$ and $G_2$ are the time intervals $1984.0-B1950.0$ and
$J2000.0-1984.0$, respectively. For details of the notions in
Equation (13) and the constructing procedure of
$\mathcal{M}_{1984.0}$, readers are referred to \citet{standish82}
and \citet{aoki83}. The numerical form of our matrix
$\mathcal{M}_{1984.0}$ shows consistency with other authors' except
the difference in the last digit. Based on the transformation matrix
we convert the positions and proper motions of 1535 FK4 bright stars
from B1950.0 to J2000.0. Figure 1 displays the position difference
of corresponding stars in the FK4 and FK5 catalogue at the epoch of
J2000.0. The top panels show the position difference in the sense of
$FK4(J2000.0)-FK5$ for right ascensions and the bottom panels for
declinations. The position differences in the southern hemisphere
are more pronounced because of the lower quality observations of the
stars. Besides, the plots show residual systematic difference versus
declinations, which imply the imperfection of the transformation
procedure from the FK4 to the FK5 reference frame. On the other
hand, the differences for most stars are less than 0.3\,arcsec in
right ascension and declination, which shows the correctness of the
transformation in the range of the error of the FK4 system.

\begin{figure}
\centering
\includegraphics[scale=.6]{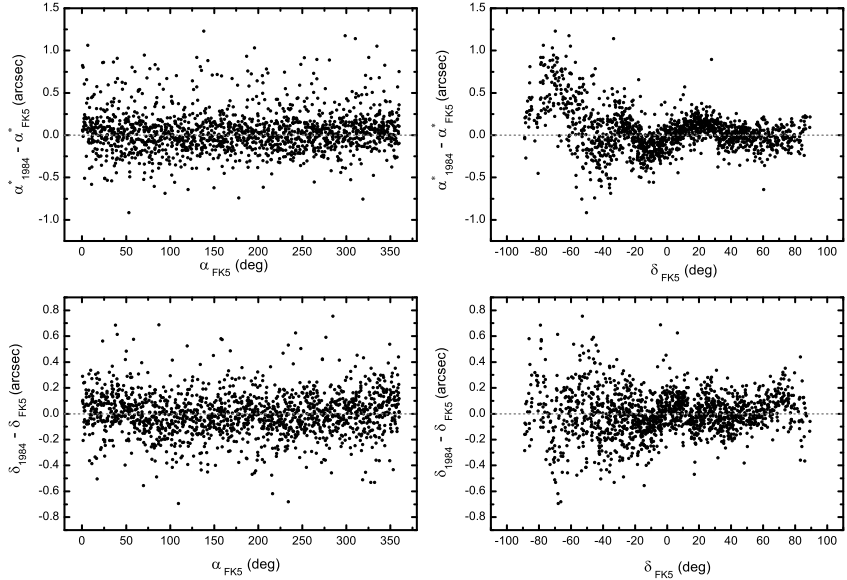}
\caption{Position differences between the FK4-based and FK5 bright
stars. The subscript $1984.0$ indicates the transformation is
operated at the epoch of 1984.0 by the matrix
$\mathcal{M}_{1984.0}$.} \label{Fig:01}
\end{figure}

\begin{figure}
\centering
\includegraphics[scale=.6]{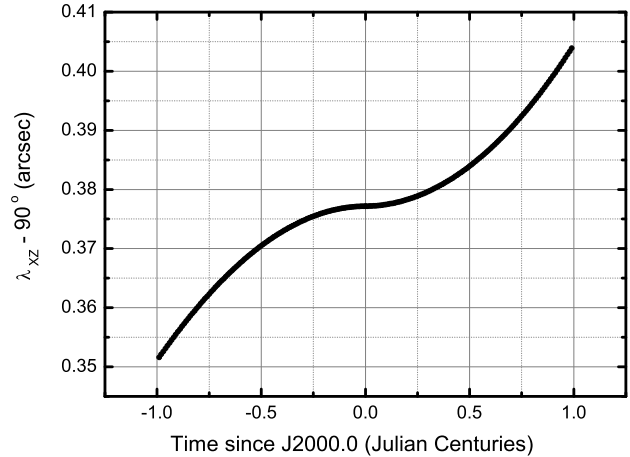}
\caption{The relationship between $\lambda_{xz}-90^{\circ}$ and time
interval since J2000.0. The symbol $\lambda_{xz}$ is the angular
separation between the $x-$ and $z-$axes, which is computed backward
and forward from J2000.0 using constant proper motion
$\mu^p_{\alpha}=0.289881$, $\mu^p_{\delta}=0.423357$,
$\mu^0_{\alpha}=-0.003708$ and $\mu^0_{\delta}=0.019808$ in
$\rm{as\,cty}^{-1}$.} \label{Fig:02}
\end{figure}

During the transformation, a correction of the time derivative of
the E-term would produce an additional effect on the proper motions
in comparison to that from the pure rotations. In fact, if we assume
that the position of an extragalactic object referred to the FK4 (or
FK5) system is fixed (because its proper motion is zero), then it
has non-zero fictitious proper motion with respect to the FK5 (or
FK4) system, by the corresponding total amount of difference in
derivative of E-term. \citet{reid04} in their study gave out the
coordinates of the north galactic pole in J2000.0 system,
$\alpha_{J}^{\,p}=12^h51^m26^s.282$, $\delta_{J}^{\,p}=+27^\circ 07'
42''.01$, and of the galactic center,
$\alpha_{J}^{\,0}=17^h45^m37^s.224$, $\delta_{J}^{\,0}=-28^\circ 56'
10''.23$, as well as the position angle of the direction of the
galactic center, $\theta_{J}=122^\circ.932$ (see their Appendix). We
tried to transform the IAU 1958 version of the definition in
Equation (1), using the rigorous method (two steps described above),
from B1950.0 to that of J2000.0 and found identical values as the
results derived by \citet{reid04}. As a byproduct, the proper
motions of both points are $\mu^p_{\alpha}=0.289881$,
$\mu^p_{\delta}=0.423357$, $\mu^0_{\alpha}=-0.003708$ and
$\mu^0_{\delta}=0.019808$ in $\rm{as\,cty}^{-1}$ which can be
attributed to the derivative of the E-term in aberration. Such a
transformation procedure would be problematic for the purpose of
calculating the orientation of the new galactic coordinate system.
Firstly, the $x-$ and $z-$ axes, pointing to the galactic center and
north galactic pole, are not perpendicular to each other with a
mismatch of $0.377\,\rm{arcsec}$ at the epoch of J2000.0. The
discrepancy will accumulate as the time interval gets longer (see
Figure 2). Secondly, the points like the intersections of the
coordinate axes and celestial sphere, and circles like the galactic
equator, which must be fixed in the FK5 system at J2000.0 are going
to move at the rate of their ``fictitious proper motions''. Figure 3
describes the distortion of the galactic plane in the FK5 reference
system if it is transformed from B1950.0, when it was a rigorous
great circle on the celestial sphere. The vectors denote the
``proper motions'' coming from the non-pure-rotation transformation.
The reduced galactic plane is a time-dependent curve. This
incoherence is attributed to the adoption of definitional values,
like in Equation (1), as the coordinates of specific objects.

On the other hand, Equation (4) is not a rigorous transfer equation
from the FK4 to the FK5 system, which would cause confusion in
observations in the galactic coordinate system. For instance,
supposing an object was observed to coincide with the galactic
center at the epoch of B1950.0 in the FK4 system ($\ell_{B}=0,\;
b_{B}=0$), its galactic coordinates cannot correspond with the
galactic center at J2000.0 ($\ell_{J}\neq 0,\;b_{J}\neq 0$), if the
equatorial position was transformed using the rigorous method from
B1950.0 to J2000.0 and then using ${\mathcal N}_{J}$ from equatorial
to galactic coordinate system at the epoch of J2000.0. In other
words, the direction $(\ell_{J}=0,b_{J}=0)$ would not correspond to
the the original object in the galactic center. Any other stars on
the celestial sphere will suffer this effect (tenth of arcsecond
level) if related to the transformation of the galactic coordinates
from the FK4 to FK5 system. Therefore the transformed galactic
coordinate system \citep{murray89} does not coincide with the
physical features as in the 1958 definitions. One may have doubt
where the actual galactic center and galactic plane are. For other
applications of the conversion, \citet{aoki86} had discussed the
optical observation and the difficulties in observing the radio
sources. According to their study, specified reference framework and
constants system are essential to avoid unnecessary confusion.

\begin{figure}
\centering
\includegraphics[scale=.60]{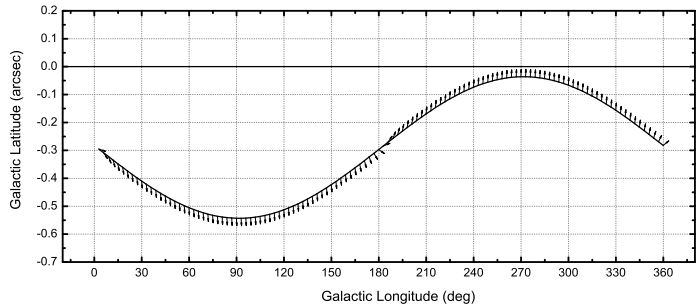}
\caption{The distortion of the galactic plane (the solid curve) in
the FK5 system at J2000.0. The vectors denote the proper motion of
certain points although they should be fixed in the FK4 system at
B1950.0. The solid line is the nominal galactic plane with
$\ell=0$.} \label{Fig:03}
\end{figure}

\subsection{Conversion of galactic coordinates from the FK5 system to the ICRS}

The establishment of the ICRS as the fundamental celestial reference
system was described by \citet{arias95}, and \citet{feissel98}. It
is a coordinate system whose origin is at the barycenter  of the
Solar system and whose axis directions are effectively defined by
the adopted 212 extragalactic radio sources observed by VLBI
technique \citep{ma98}. Although, more recently, the ICRF-2
\citep{bob10} which includes 3414 extragalactic radio sources was
put into effect on January 1, 2010, the definition of the ICRS was
not changed. The orientation of the ICRS was designed to be
consistent with the dynamical equator and equinox (realized by the
FK5 catalogue) at the epoch of J2000.0.

Systematically, a \emph{frame bias matrix} $\mathcal B$ is needed to
convert ICRS data to the dynamical equator and equinox of J2000.0.
It is a fixed set of very small rotations and can be used as
follows:
\begin{equation}
{\bf{A}}_{J} = \mathcal B\, {\bf{A}}_{ICRS}
\end{equation}
where the column vector ${\bf A}$ has the similar form as before
except for the subscripts which denote the benchmark reference
system or epoch. The matrix $\mathcal B$ is, to the first order
\citep{mccarthy03},
\begin{equation}
\mathcal B = \left(
\begin{array}{ccc}
1 & \rm{d}\alpha & -\xi_0 \\
-\rm{d}\alpha & 1 & -\eta_0\\
\xi_0 & \eta_0 & 1
\end{array}
\right)
\end{equation}
where the angles $\eta_0=-6.819\pm0.0015 \,\rm{mas}$ and
$\xi_0=-16.6171\pm0.0080\,\rm{mas}$ are the ICRS pole offsets in
$x$, $y$ directions, and ${\rm{d}}\alpha=14.6\pm 0.5 \,\rm{mas}$ is
the offset in the equinox between the ICRS and J2000.0 dynamical
system \citep{chapront02}. See \citet{hilton04} for detailed
description. To the second order, the more precise format is
\begin{equation}
\mathcal B= \left(
\begin{array}{ccc}
1-\left(\rm{d}\alpha_0^2+\xi_0^2\right)/2 & \rm{d}\alpha_0 & -\xi_0\\
-\rm{d}\alpha_0-\eta_0 \xi_0 & 1-\left(\rm{d}\alpha_0^2+\eta_0^2\right)/2 & -\eta_0\\
\xi_0-\eta_0\rm{d}\alpha_0 & \eta_0+\xi_0\rm{d}\alpha_0 &
1-\left(\eta_0^2+\xi_0^2\right)/2
\end{array}
\right).
\end{equation}
The above matrix can also be used to transform vectors from ICRS to
the FK5 system at the epoch of J2000.0 by simply substituting
$\rm{d}\alpha_0=-22.9$, $\eta_0=-19.9$ and $\xi_0=9.1\,\rm{mas}$
\citep{mignard97}.

For the galactic coordinate system, we can use Murray's method for
the purpose of constructing the equatorial-galactic transformation
matrix in the ICRS. Neglecting the relative rotation between the FK5
system and the ICRS and assuming the axes of the galactic coordinate
system are absolute, we get the relation between ${\mathcal N}_{J}$
and $\mathcal {N}_{ICRS}$:
\begin{equation}
{\mathcal N}_{ICRS}={\mathcal N}_{J}\,{\mathcal B}.
\end{equation}
The numerical form of the matrix $\mathcal {N}_{ICRS}$ is:
\begin{equation}
\mathcal{N}_{ICRS}= \left(
\begin{array}{rrr}
-0.054875657707 &    -0.873437051953 &    -0.483835073621  \\
+0.494109437203 &    -0.444829721222 &    +0.746982183981  \\
-0.867666137554 &    -0.198076337284 &    +0.455983813693
\end{array}
\right)
\end{equation}
by adopting $\mathcal B$ as in Equation (16). The direction of the
galactic pole and the position angle in the ICRS should be
\begin{eqnarray}
&&\alpha_{ICRS}^{\,p}  = \;\;12\,^h51\,^m26\,^s.27469  \nonumber  \\
&&\delta_{ICRS}^{\,p}  \,=  +27\,^\circ07\,'41\,''.7087  \\
&&\theta_{ICRS}  = \;\;122\,^\circ.93192526 .    \nonumber
\end{eqnarray}
The difference between Equation (10) and (19) is at the order of
several milli-arcseconds. The above discussion only considered the
orientation bias between the FK5 and the ICRS. In the sense of
inertia, the galactic coordinates have a slow spin as well as
regional distortion with respect to the ICRS due to the
imperfections in the FK5 system. Such a derived galactic coordinate
system is still an expediency because it is not fixed in the ICRS.

In order to reveal the systematic difference between the FK5 and
ICRS, which is the optically realized by the \emph{Hipparcos
catalogue}, various authors \citep{zhu99,mignard00,zhu00,walter05}
have intensively compared the proper motion systems of the two
catalogues. To summarize the studies, the global rotation between
the FK5 and ICRS, which is derived from the comparison of the proper
motion systems of FK5 and \emph{Hipparcos} show obvious
inconsistencies (approximately -1.3\,$\rm{mas\,yr}^{-1}$  by
rotational offsets of the \emph{Hipparcos} and FK5 proper motion
systems), compared to that predicted by the precession correction,
$\Delta p$, and correction of the fictitious motion of the equinox,
$\Delta e +\Delta \lambda$, from the unprecedented-accurate VLBI and
LLR observations \citep{charlot95,capitaine04}. The mismatch is
probably caused by the error and nonrigid nature of the FK5
proper-motion system and this strongly implies the existence of
non-inertiality of the FK5 system. Directly speaking, the FK5 is
rotating with respect to the distant radio sources even if the
precession correction is applied to the IAU (1976) precession
constants. Furthermore, the region-, magnitude-, and color-dependent
systematic errors between the FK5 and \emph{Hipparcos} prevent us
from carrying out a perfect alignment of the FK5 system with the
ICRS by using simple rotation matrices. \citet{soma00} analyzed
Lunar occultations and found much larger errors in the rotation of
the \emph{Hipparcos} reference frame than those given by
\emph{Hipparcos} team, which suggested the possibility of a
non-inertiality of the \emph{Hipparcos} system. Therefore we come to
the conclusion that the galactic coordinate system derived from a
transformation is neither space-fixed nor time-independent. The
system is rotating with respect to the distant extragalactic objects
even if it is exactly fixed to the FK5 system.

\section{Reconsidering the definition of the galactic coordinate system}
\label{sect:reconsider}

We note that the present galactic coordinate system does not meet
the defining criterion of the fundamental reference system, namely,
non-rotating with respect to the ensemble of distant extragalactic
objects. Using the transformation method we can not tie the galactic
coordinate system, which was originally defined in 1958, uniquely to
the ICRS. There must be residual rotation (slow spin) due to the
non-inertial property of the FK5 system which is unacceptable for
high-precision observations at the level of one micro-arcsecond.
There actually exist several different galactic coordinate system,
on the basis of the FK4, the FK5, or the ICRS, having orientation
offsets and relative rotation among one another. The least (relative
rotation neglected) differences (between Equation (19) and (10)) in
the principal values that define the galactic coordinate system in
the sense of $ICRS-FK5$ are roughly:
\begin{eqnarray*}
&&\Delta \alpha^{\,p}=-1.2 \,\rm{mas} \\
&&\Delta \delta^{\,p}=-4.4 \,\rm{mas} \\
&&\Delta \theta=+24.1 \,\rm{mas}.
\end{eqnarray*}
These offsets are much larger than the axes stability of the ICRS
($\simeq20\rm{\mu as}$), so the old definition is not appropriate in
current fundamental reference system. In order to avoid the
confusion in choosing the galactic coordinate system on one hand and
to clarify the concept of the galactic coordinate system on the
other, we think that a new definition is needed. Since more and more
important astrometric data, such as UCAC3 \citep{zacharias10}, PPMX
\citep{roser08} and PPMXL \citep{roser10} are presented in the ICRS
framework, the galactic coordinate system should be directly
connected with the ICRS for practical observations and studies of
the galactic structure, kinematics and dynamics, especially for the
galactic center.

To successfully achieve the objective of establishing a galactic
coordinate system fixed to the ICRS, which was originally defined in
the radio bandpass, the principal plane and the galactic center
should likewise be observed using radio techniques. For example,
direct observation of the compact radio source Sagittarius A* (Sgr
A*) at the dynamical center of the Milky Way \citep{reid09} might
give the starting point of the galactic longitude. With measurements
using the Very Long Baseline Array (VLBA), the best estimation of
the absolute position (J2000.0) of Sgr A* \citep{reid04} is
\begin{eqnarray}
&& \alpha=17\,^h45\,^m40\,^s.0400 \nonumber \\
&& \delta=-29^\circ00'28''.138,
\end{eqnarray}
and the proper motion of Sgr A* is: $\mu_\ell = -6.379\pm0.026
\,\rm{mas\,yr}^{-1}$ and $\mu_b=-0.202\pm0.019 \,\rm{mas\,yr}^{-1}$
that may reflect the galactic rotation and the peculiar motion of
the Sun. The uncertainty of the position ($\pm 10$\,mas) is
dominated by the position of J1745-2820. Considering that the proper
motion of Sgr A* is nonzero, the observational position should be
reduced to the fixed epoch J2000.0. Suppose that Sgr A*, whose
position in the ICRS is given by Equation (20), is at the galactic
center (i.e. the Sgr A* is regarded as the zero point of the
system), we can find the position of the corresponding galactic pole
${\bf A}_{New}^p$ from following equations:
\begin{eqnarray}
&& {\bf A}_{New}^{\,p} \cdot {\bf A}_{New}^0=0 \nonumber \\
&& {\bf A}_{New}^{\,p} \cdot \left( {\bf A}_{New}^0 \times {\bf
A}_{ICRS}^p \right)=0,
\end{eqnarray}
where ${\bf A}_{ICRS}^p$ is the unit column vector along the
galactic pole derived from Equation (19) and ${\bf A}_{New}^0$ the
position vector of the new galactic center defined by Equation (20).
This new galactic pole is located closest to the old one (${\bf
A}_{ICRS}^p$). The resulting coordinates of the new galactic pole in
the ICRS are (printed to one micro-arcsecond accuracy):
\begin{eqnarray}
&&\alpha_{New}^{\,p}  = \;\;12\,^h51\,^m36\,^s.7151981  \nonumber  \\
&&\delta_{New}^{\,p}  \,=  +27\,^\circ06\,'11\,''.193172.
\end{eqnarray}
Accepting the coordinates in Equation (20) as a new definition of
the zero point of the galactic coordinate system, and the position
described in Equation (22) as the the position of the new galactic
pole, we obtain the transformation matrix:
\begin{equation}
\mathcal{N}_{New}= \left(
\begin{array}{rrr}
-0.054657353964 &    -0.872844082054 &    -0.484928636070  \\
+0.494055821648 &    -0.445679169947 &    +0.746511167077  \\
-0.867710446378 &    -0.198779490637 &    +0.455593344276
\end{array}
\right),
\end{equation}
as well as the position angle:
\begin{equation}
\theta_{New}=123^{\circ}.0075021536.
\end{equation}
These values are not consistent with those in previous sections
because we adopted different coordinates for the galactic center
which have several arcminutes disagreement.

As for the formal definition of the galactic reference system in the
ICRS, re-observation of the galactic center and the HI principlal
plane in the radio bandpass is necessary and multi-wavelength
observations may be needed for practical application purposes.
Because the height of the Sun above the galactic plane $z_0$ is not
zero, the equatorial plane of the galactic coordinate system, which
should theoretically pass through the Sun and the galactic center,
should be set as close to the H\textrm{I} principlal plane as
possible. Then the obtained galactic orientation parameters should
be regarded as the definition of the new galactic coordinate system.
Since the galactic coordinate system is fixed in the ICRS, there is
no epoch associated with a galactic coordinate, so it is technically
incorrect to say that the galactic coordinate system is a ``J2000.0
system''.

As the newly defined galactic coordinate system should become
effective, the data in old (FK5) galactic coordinate system could be
calculated at the epoch of J2000.0 by multiplying the bias matrix
$\mathcal B$ where high accuracy is not required. In the optical
frequency range, the new galactic coordinate system should be fixed
in the \emph{Hipparcos} celestial reference system (HCRS) which is
aligned to the ICRS using various methods \citep{kovalevsky97}. The
estimated uncertainty of the link corresponds to a standard error of
0.6\,mas in axes alignment and 0.25\,$\rm{mas\,yr}^{-1}$ in the rate
of rotation.

The galactic coordinate system that we considered is a convenient
barycentric system. If we know the accurate distance of the Sun out
of the galactic plane, one could consider rotating the system to
that which would be observed by an observer exactly in the plane.
This would be approximately a 0.1 degree (the angle of the Sun with
respect to to the galactic plane viewed from the galactic center)
rotation so that the origin of the reference system is certainly in
the galactic plane. \citet{kovalevsky03} pointed out that a
galactocentric rather than a barycentric reference frame should be
used because upcoming space astrometric missions Gaia \citep{esa00}
and SIM \citep{danner99} will drastically improve the astrometric
accuracy to micro-arcsenond level. If one wishes to define a
reference system centered at the galactic center, one need to know
the exact galactic constants like the galactocentric distance of the
Sun $(R_0)$, Oort constants $(A,\,B)$, the vertical distance of the
Sun above the galactic plane $(z_0)$ and any other parameters that
define the position and the motion of the Sun. However all these
values are estimated and still open for research; there is no
official resolution about the definite values of galactic constants
--- different galactic constants would lead to different reference
systems. The definition of a galactocentric reference system is at
risk and uneconomical even if the galactic constants are authorized
because all observations were made in a barycentric system. The
astrometric data (positions, proper motions and parallaxes of
celestial bodies) must be converted from barycentric to
Galactocentric reference system by some more complicated
transformation which are related to the galactic constants, rather
than ordinary rotation once such coordinate system is defined.

\section{Conclusion}
\label{sect:conclusion}

We have reviewed the history of the galactic coordinate system,
which was primitively defined by the International Astronomical
Union in 1958. The efforts of transforming the galactic coordinate
system from the FK4 to the FK5, and from the FK5 to the currently
used fundamental reference system -- ICRS, have led to some
confusions and misunderstandings. The galactic coordinate system is
rotating with respect to the ICRS and the effect can not be
eliminated by simple rotation due to the complexity of the
fundamental catalogues (FK4, FK5). As a temporary expedient, we have
derived the rotation matrix from the equatorial to the galactic
coordinate system in the framework of ICRS using the bias matrix,
and this can be used only when high accuracy is not needed. Our
proposal for the new definition of the galactic coordinate system
makes use of current observation of the supermassive black hole Sgr
A* and is, of course, still open for discussion. With the
development of radio astronomy to unprecedented accuracy, we suggest
that the galactic coordinate system be directly connected to the
ICRS by observing the galactic center and plane. Therefore, it is
recommended that the definition of the galactic coordinate system be
reconsidered by the relevant IAU Commissions.

\begin{acknowledgements}
This work was funded by the National Natural Science Foundation of
China (NSFC) under No. 10973009. The authors wish to thank Drs. L.
Chen, D.L. Kong, and Prof. M. Miyamoto for their interest and
encouragement. We are grateful to Prof. W.J. Jin, M. Reid, Dr. E.
H{\o}g and in particular, to M. S\^{o}ma for their valuable
comments, suggestions and discussion. Some points are revised
according to their suggestions. We are also indebted to Prof. W.F.
van Altena for his very careful correction of the language.
\end{acknowledgements}

\end{document}